\documentclass[aps,preprint,tightenlines,groupedaddress,nofootinbib]{revtex4}

\usepackage{graphicx} 

\newcommand{\FSA}{0.5\textwidth}
\newcommand{\FSB}{\textwidth}
\newcommand{\vecr}{\mbox{\boldmath $r$}}

\newcommand{\sigmab}{\bar{\sigma}}

\newcommand{\betwo}{$B(E2, 0^{+}_1 \rightarrow 2^{+}_1)$ }
\newcommand{\twop}{$2^{+}$ }

\begin{document}

\title{
Pairing effects on the collectivity of quadrupole states around $^{32}$Mg
}

\author{M. Yamagami,$^{1}$ Nguyen Van Giai$^{2}$\\
{\it \small $^{1}$Department of Physics, Graduate School of Science,
Kyoto University,}\\
{\it \small Kyoto 606-8502, Japan}\\
{\it \small $^{2}$Institut de Physique Nucl\'eaire, IN$_{2}$P$_{3}$-CNRS,
91406 Orsay Cedex, France} }

\begin{abstract}
The first 2$^{+}$ states in N=20 isotones including neutron-rich nuclei 
$^{32}$Mg and $^{30}$Ne are studied by 
the Hartree-Fock-Bogoliubov plus
quasiparticle random phase approximation method based on the Green's function 
approach. The residual interaction between the quasiparticles is consistently 
derived from the hamiltonian density of Skyrme interactions with explicit 
velocity dependence. The \betwo transition probabilities and the excitation energies 
of the first 2$^{+}$ states are well described within a single framework. 
We conclude that pairing effects account largely for the anomalously large B(E2) 
value and the very low excitation energy in $^{32}$Mg.
\end{abstract}

\maketitle

\section{Introduction}

The pioneering observation in 1975 of the anomalous binding energy gain 
in very neutron rich Na isotopes revealed the breaking of the N=20 shell 
closure and the 
possibility of deformation \cite{TK75}.
The evidences of the breaking of N=20 shell closure in neutron-rich Mg and Ne isotopes 
are more clearly seen from the observations of E2 properties, the large B(E2) value 
in $^{32}$Mg \cite{MI95} and the low excitation energies of the first \twop states 
in $^{32}$Mg \cite{GD84} and $^{30}$Ne \cite{YN02}.

Several theoretical studies have been done to describe the anomalous
binding energy and E2 properties in neutron-rich nuclei around N=20.
Constrained Hartree-Fock (HF) calculations of Na isotopes \cite{CF75}
have been performed
and $^{31}$Na was suggested as deformed. Early studies made by Wildenthal and Chung
showed that shell model calculations within the $sd$ shell model space cannot
explain the extra binding energies in this region. Subsequent
shell model calculations \cite{WC80,WS81,PR87,WB90} have
demonstrated that the inclusion of the $fp$ shell into the shell model active space is essential.
The effects of the breaking of the N=20 shell closure are clearly shown in the 
description of the B(E2) values and the excitation energies in $^{32}$Mg and $^{30}$Ne 
\cite{FO92,CN98,UO99}. The neutron 2p-2h configurations across the N=20 shell 
imply deformation of these nuclei. 
However, in the framework of the mean-field 
approximation with pairing correlations, like Skyrme Hartree-Fock-Bogoliubov (HFB) 
calculations, the calculated ground states in $^{32}$Mg and $^{30}$Ne 
turn out to be spherical 
(see, e.g., \cite{TF97,RD99}). One possible way to describe the (dynamical) 
deformation is to include correlations beyond the mean field.
Generator coordinate method (GCM)\cite{PG00,RE02,RE03} 
and antisymmetrized molecular dynamics calculations \cite{KH02}
have been done in this direction.

Nevertheless, the experimental evidence of deformation in 
$^{32}$Mg is not well established. 
The energy ratios of the first 4$^{+}$ and \twop states, $E(4^+)/E(2^+)$, are 3.0 
in $^{24}$Mg and 3.2 in $^{34}$Mg \cite{YS01}, and these values are undoubtedly close 
to the rigid rotor limit of 3.3. 
On the other hand, the ratio is 2.6 in $^{32}$Mg, and this value is 
in between the rigid rotor limit and the harmonic vibration limit 2.0 
\cite{YS01,Gu02}. Moreover, the B(E2) value (in single-particle units) is 
15.0$\pm$2.5 in $^{32}$Mg. This value is larger than in the other stable N=20 
isotones but smaller than in other deformed Mg isotopes 
(21.0$\pm$5.8 in $^{24}$Mg, and 19.2$\pm$3.8 in $^{34}$Mg).

Generally speaking, the neutron 2p-2h configurations can originate
not only from 
deformation effects but also from neutron pairing correlations.
In the $^{32}$Mg nucleus these two effects may coexist and make the large B(E2) 
value and the low excitation energy of the \twop state. 
In shell model studies it is not clear 
which effect is more essential to describe these anomalous properties. 

Recent angular-momentum projected GCM calculations with the 
Gogny force \cite{RE02,RE03} 
were successful to reproduce the systematic trend of the B(E2) values and the excitation 
energies of the first \twop and 4$^{+}$ states in Mg and Ne isotopes. 
However, the calculated excitation energies of the first \twop states 
are somewhat higher 
than the experimental data (e.g., about 1.5 MeV in $^{32}$Mg and about 2.1 MeV 
in $^{30}$Ne). 
These discrepancies may be explained by the weakness of neutron pairing 
correlations around the spherical ground states in $^{32}$Mg and $^{30}$Ne 
in the corresponding HFB mean fields. 

The purpose of this paper is to emphasize how neutron pairing correlations
play an essential role in the description of E2 properties in $^{32}$Mg and
$^{30}$Ne. The existence of neutron pairing correlations means the breaking of the
N=20 shell closure. As we will see, the appearance of neutron pairing
correlations is related to a special mechanism  in loosely bound systems.
We study the first \twop states in N=20 isotones
in the framework of self-consistent
quasiparticle random phase approximation (QRPA)
with Skyrme interactions \cite{YK03}.
The QRPA equations are solved in coordinate space by using the Green's function
method \cite{KS02}.
Spherical symmetry is assumed for simplicity.
The residual interaction between the quasiparticles is self-consistently
derived from the hamiltonian density of Skyrme interaction that has an explicit
velocity dependence.
We will show that the B(E2) values and the excitation energies of the first \twop
states in N=20 isotones, from the stable nucleus $^{38}$Ar to the neutron-rich nuclei
$^{32}$Mg and $^{30}$Ne are well described within a single framework and
a fixed parameter set. The paper is organized as follows. In Sect. \ref{SEC-QRPA}
we briefly describe the HFB plus QRPA calculations that we have done. In Sect.
\ref{SEC-GS} we present the general results for the ground states of the N=20 isotones
studied here. In Sect. \ref{SEC-BE2} we discuss the calculated and experimental $E2$
properties of these nuclei. Conclusions are drawn in Sect. \ref{SEC-CONC}.


\section{HFB-QRPA calculations} \label{SEC-QRPA}


\subsection{Formulation}

We use the approach of self-consistent HFB-QRPA calculations with
Skyrme interactions \cite{YK03,KS02}. By self-consistent we mean that the HFB
mean fields are determined self-consistently from an effective force 
and the residual interaction of the QRPA problem is derived from the same
force. 
The QRPA problem is solved by the response function method 
in coordinate space. A detailed account of the method can be found in
Ref.\cite{KS02}. Here, we just recall the main steps of the calculation.
The QRPA Green's function ${\bf G}$ is solution of a Bethe-Salpeter equation,
\begin{equation}
{\bf G} = {\bf G}_0 +{\bf G}_0 {\bf V}{\bf G}~.
\end{equation}
The knowledge of ${\bf G}$ allows one to construct the response function of
the system to a general external field, and the strength distribution of the
transition operator corresponding to the chosen field is just proportional
to the imaginary part of the response function.

In Eq.(1) 
the unperturbed Green's function ${\bf G}_0$ is defined as
\begin{eqnarray}
G_0^{\alpha\beta}(\vecr\sigma,\vecr'\sigma';\omega) &=& 
\sum_{i,j}
\frac{ W_{i,j}^{\alpha 1}(\vecr\sigma)
        [ W_{i,j}^{\beta 1*}(\vecr'\sigma') ]_{-} }
{\hbar\omega -(E_{i} + E_{j}) + i\eta}  
- 
\frac{ W_{i,j}^{\alpha 2*}(\vecr\sigma)
        [ W_{i,j}^{\beta 2}(\vecr'\sigma')  ]_{-} }
{\hbar\omega +(E_{i} + E_{j}) + i\eta}~,
\label{EQGF}
\end{eqnarray}
where the functions $W(\vecr \sigma)$ are introduced as
\begin{eqnarray}
W_{i,j} (\vecr\sigma)=\left(
\begin{array}{cc}
U_{i} (\vecr\sigma) V_{j} (\vecr\sigma) 
  & V_{i} (\vecr,\sigma) U_{j} (\vecr\sigma) \\
U_{i} (\vecr\sigma) U_{j} (\vecr\sigmab) 
 & V_{i} (\vecr\sigma) V_{j} (\vecr\sigmab) \\
-V_{i} (\vecr\sigma) V_{j} (\vecr\sigmab) 
 & -U_{i} (\vecr\sigma) U_{j} (\vecr\sigmab) \\
\end{array} \right)~.
\label{EQDFW}
\end{eqnarray}
Here, the $U(\vecr), V(\vecr)$ are quasiparticle wave functions,
the index $\alpha$ ($\alpha$=1,2,3) stands for particle-hole (ph), particle-particle (pp) 
and hole-hole (hh) channels.
The notation 
$f(\vecr\sigmab)\equiv -2\sigma f(\vecr-\sigma)$  
indicates time-reversal
and $\left[ W_{i,j} \right]_{-} = W_{i,j} - W_{j,i}$.

The residual interaction ${\bf V}$ between quasiparticles is derived from the 
Hamiltonian density $<H>$ of Skyrme interaction by the so-called Landau procedure,
\begin{equation}
V_{\alpha \beta} \left( {\vecr\sigma \tau ,\vecr'\sigma '\tau '} \right) 
= \frac{{\partial ^2 <H> }}{{\partial \rho _\beta  
\left( {\vecr'\sigma '\tau '} \right)\partial \rho _{\bar \alpha } 
\left( {\vecr\sigma \tau } \right)}}~. \label{eqLM}
\end{equation}
The notation ${\bar \alpha }$ means that 
whenever $\alpha$ is pp (hh) then ${\bar \alpha }$ is hh (pp).
The normal and abnormal densities are defined as 
\begin{eqnarray}
\left(
  \begin{array}{c}
   \rho_{ph}(\vecr\sigma) \\
   \rho_{pp}(\vecr\sigma) \\
   \rho_{hh}(\vecr\sigma)
  \end{array}
\right)
=
\left(
  \begin{array}{c}
   \rho(\vecr\sigma) \\
   \kappa(\vecr\sigma) \\
   \bar{\kappa}(\vecr\sigma)
  \end{array}
\right)
=
\left(
  \begin{array}{c}
   <0|\psi^{\dagger}(\vecr\sigma)\psi(\vecr\sigma)|0> \\
   <0|\psi(\vecr\sigmab)\psi(\vecr\sigma)|0> \\
   <0|\psi^{\dagger}(\vecr\sigma) \psi^{\dagger}(\vecr\sigmab)|0> 
  \end{array}
\right)~.
\end{eqnarray}
The residual interaction ${\bf V}$ has an explicit momentum dependence, 
\begin{equation}
{\bf V}(\vecr,\vecr')={\bf F}[\overleftarrow{\Delta}_U +\overleftarrow{\Delta}_V, 
\overrightarrow{\Delta}_U +\overrightarrow{\Delta}_V,
\overleftarrow{\nabla}_U \pm \overleftarrow{\nabla}_V,
\overrightarrow{\nabla}_U \pm \overrightarrow{\nabla}_V]
\delta (\vecr-\vecr'). \label{EQ_VRESI}
\end{equation}
The explicit form of the form factor ${\bf F}$ is shown in Ref.\cite{YK03}.
The operators with $\leftarrow$ ($\rightarrow$) act on 
the coordinate $\vecr$ ($\vecr'$), 
and the operators with the index $U$ ($V$) operate on 
the quasiparticle wave functions 
$U(\vecr)$ ($V(\vecr)$) only. These momentum dependence are explicitly treated in our calculation.
Because we calculate only natural parity (non spin-flip) excitations, 
we drop the spin-spin part of the residual interaction. The Coulomb and spin-orbit 
residual interactions are also dropped.

We have studied the influence of these momentum dependent terms and we have
found that they can be important\cite{YK03}. 
In a fully consistent calculation the spurious center-of-mass 
state should come out at zero energy. In practice, the full velocity
dependence of the residual interaction (\ref{EQ_VRESI}) is replaced by a
Landau-Migdal form 
to reduce the computational efforts\cite{KS02}. Then,  
the self-consistency between the mean-field and the residual interaction is 
broken. To recover the self-consistency approximately, the residual 
interaction has to be  
renormalized by a factor adjusted so as to have the spurious state at zero
energy. For example, the renormalization in Ref.\cite{KS02} is about 20\%. 
If the residual interaction in Eq.(\ref{EQ_VRESI}) is used, 
the renormalization factor is very close to 1 and the self-consistency is well 
recovered. For comparison, if one drops all momentum-dependent terms in
Eq.(\ref{EQ_VRESI}) the renormalization factor would be about 0.6 .

Another important aspect is related to the description of the low-lying states.
The B(E$\lambda$) transition probability 
is very sensitive to the treatment of the residual interaction. 
For example, the \betwo in $^{20}$O calculated with SkM* parameter 
with (without) these momentum dependent terms is 34.1 (20.9) $e^2 fm^4$. 
The B(E2) value increases by 64 \% \cite{YK03}. 
The experimental data is $28\pm 2$ $e^2 fm^4$ \cite{RM87} which is 
close to the value calculated with the momentum dependent terms. 
Thus, the full residual interaction (\ref{EQ_VRESI}) is important for
describing quantitatively 
the low-lying 
states and for comparison with experimental data. 


\subsection{Inputs}

We apply the above formalism to study the first \twop states in N=20 isotones,
$^{30}$Ne, $^{32}$Mg, $^{34}$Si, $^{36}$S and $^{38}$Ar. 
The ground states are given by Skyrme-HFB calculations.
The HFB equation is diagonalized on a Skyrme-HF basis calculated 
in coordinate space with a box boundary condition \cite{GB94,TH95,YM01}. 
Spherical symmetry is imposed on quasiparticle wave functions.
The quasiparticle cut-off energy is taken to be   
$E_{cut}=50$ MeV, and the angular momentum cut-off is $l_{max}=7\hbar$ 
in our HFB and QRPA calculations.

\begin{figure}
  \begin{center}
    \includegraphics[width=\FSA,clip]{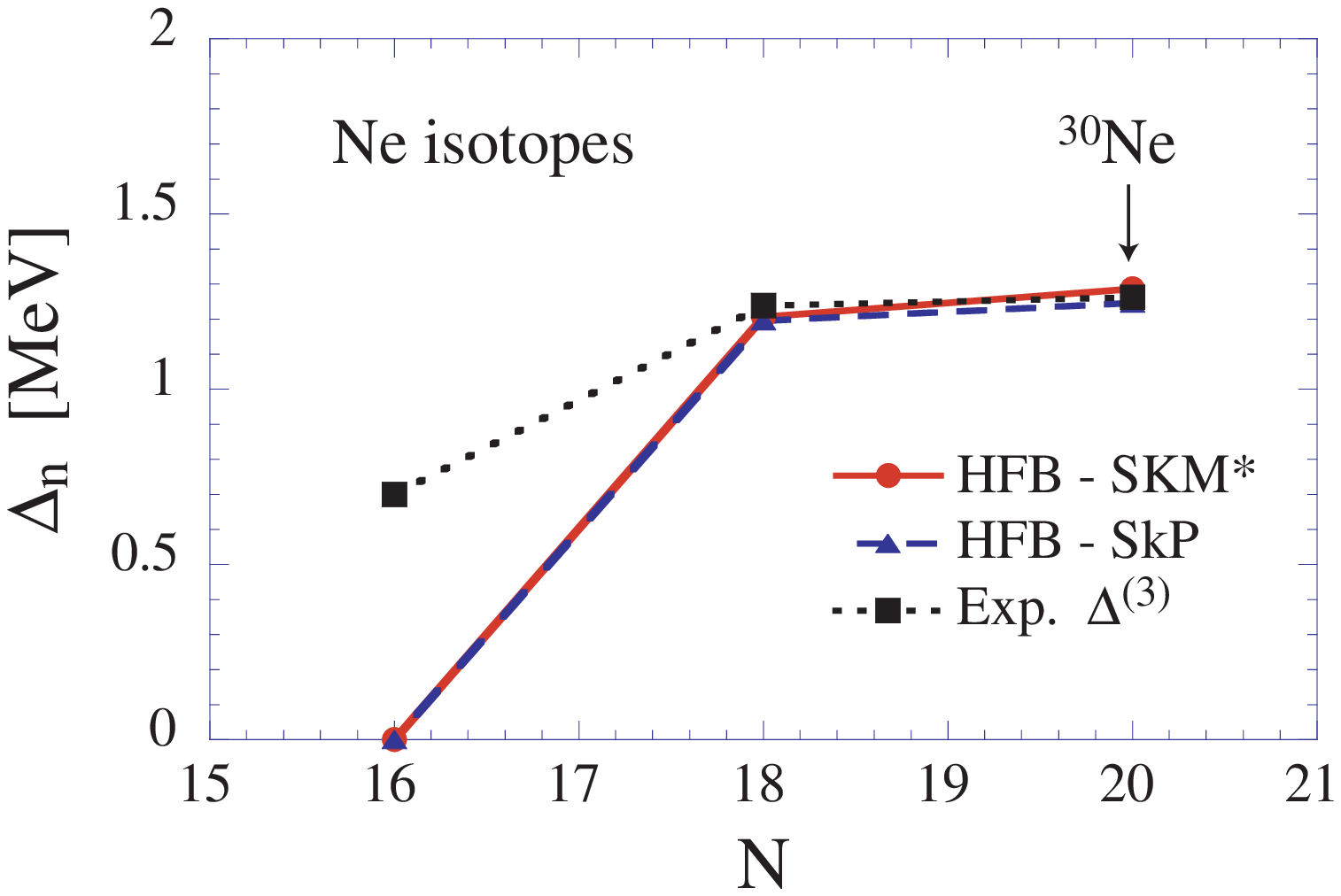}
  \end{center}
  \caption{HFB neutron pairing gaps in $^{26,28,30}$Ne 
   calculated with SkM* and SkP.
   The pairing strengths $V_{pair}$ are fixed so as to reproduce 
   the experimental neutron gap in $^{30}$Ne.
   The experimental pairing gaps are extracted by using 
   the 3-point mass difference formula \cite{SD98}.}
  \label{FIG_Ne_DEL}
\end{figure}

The Skyrme parameter sets SkM*\cite{BQ82} and SkP\cite{DF84}
are used for the HF mean-field, and the density-dependent,
zero-range pairing interaction 
\begin{eqnarray}
V_{pair} \left( \vecr,\vecr' \right) = V_{pair} 
\left[ 1 - \left( \frac{\rho \left( \vecr \right)} {\rho_c}  \right)^\alpha  \right]
\delta \left( \vecr - \vecr' \right),
\end{eqnarray}
is adopted for the pairing field.  
The parameters $\alpha$ and $\rho_c$ are fixed as
$\alpha=1$ and $\rho_c=0.16$ fm$^{-3}$. The strength $V_{pair}$ is determined so
as  to reproduce the experimental neutron pairing gap in $^{30}$Ne, 
$\Delta_{n, exp.}(^{30}\mbox{Ne})=1.26$ MeV.  
$^{30}$Ne is the lightest mass even-even N=20 nucleus. 
The experimental pairing gaps are extracted by using 
the 3-point mass difference formula\cite{SD98},
$\Delta_n(N) = \Delta_n^{(3)}(N-1)=\frac{(-1)^{N}}{2} [E(N-2)+E(N)-2E(N-1)]$.
On the other hand, the average pairing gap in HFB calculations  
is defined as the integral of the pairing field, 
$\bar{\Delta}_n = \int d\vec{r} \tilde{\rho}_n (\vec{r}) \Delta_n(\vec{r})/
\int d\vec{r} \tilde{\rho}_n (\vec{r})$ \cite{Ma01}. 
The pairing strength adopted for SkM* is $V_{pair}=-418$ MeV fm$^{-3}$,   
and for SkP, $V_{pair}=-400$ MeV fm$^{-3}$.
Fig.\ref{FIG_Ne_DEL} shows the experimental and the calculated pairing gaps 
in $^{26,28,30}$Ne. With these Skyrme parameters and pairing strengths, 
we get finite pairing gap in $^{30}$Ne (vanishing of N=20 shell gap) 
and zero pairing gap in $^{26}$Ne (appearance of N=16 shell gap) at the same time.


\section{Ground state properties}  \label{SEC-GS}

\begin{figure}
  \begin{center}
    \includegraphics[width=\FSA,clip]{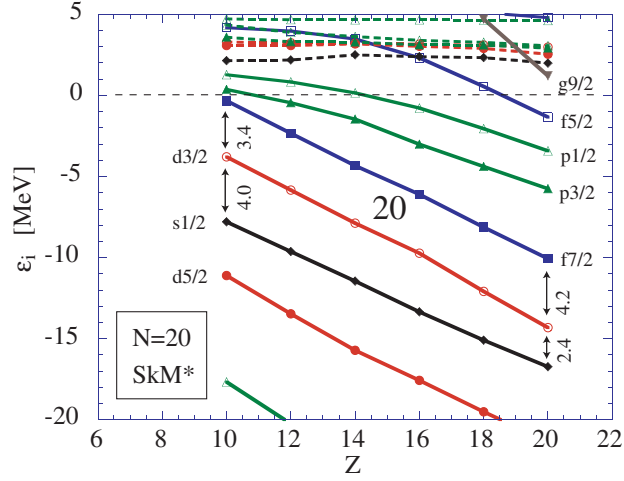}
  \end{center}
  \caption{HF neutron single-particle levels in N=20 isotones 
           calculated with SkM*. Solid lines correspond to bound and
   resonance-like states, dashed lines to positive energy
   discretized states. }
  \label{FIG_SPL}
\end{figure}

Fig.\ref{FIG_SPL} shows the neutron single-particle levels 
in N=20 isotones calculated in HF with the SkM* force.
Results with the SkP force are qualitatively the same.
The N=20 shell gaps change from 4.2 MeV in $^{40}$Ca to 3.4 MeV in $^{30}$Ne.
The N=16 shell gaps change from 2.4 MeV in $^{40}$Ca to 4.0 MeV in $^{30}$Ne.
Within HF we can describe the vanishing of N=20 magicity and the  
appearance of N=16 magic number at the same time.
The change of N=20 shell gaps looks moderate in comparison with the effective 
single-particle energies in shell model calculations\cite{UO99}.
However, the definitions are different. In shell model calculations 
the single-particle 
energies are inputs of calculations and they are determined so as to reproduce 
the neutron separation energies and the one-particle spectra of $^{17}$O and 
$^{41}$Ca, and the change of the effective single-particle energies according to 
proton number are due to the change of the many-body correlations. 
On the other hand, the change of the single-particle energies in mean-field 
calculations reflects the self-consistent change of the mean-field potential.

An important feature in Fig.\ref{FIG_SPL} is the behavior of low-{\it l} orbits,
$2p_{3/2}$ and $2p_{1/2}$ in the $fp$ shell.
As the proton number decreases the single-particle energies of
the high-{\it l} orbit
$1f_{7/2}$ change almost linearly while the changes of
$2p_{3/2}$ and $2p_{1/2}$ energies
become very slow around zero energy. Moreover, the spin-orbit splitting
of $2p_{3/2}$ and $2p_{1/2}$ states becomes smaller.
As pointed out by Hamamoto {\it et al.}\cite{HL01} these effects can be
understood by different {\it l}-dependences of the kinetic energy and the spin-orbit
form factor as the single-particle energy comes close to zero.
Because of these different {\it l}-dependences of the single-particle energies,
the level density in the $fp$ shell becomes higher with decreasing proton
number,  
and the three orbits $1f_{7/2}$, $2p_{3/2}$ and $2p_{1/2}$ 
become almost degenerate in $^{30}$Ne.
We can describe this behavior naturally by solving the HF and HFB
equations 
in coordinate space but it is difficult to get this property by the methods  
based on the harmonic oscillator basis.

\begin{figure}
  \begin{center}
    \includegraphics[width=\FSB,clip]{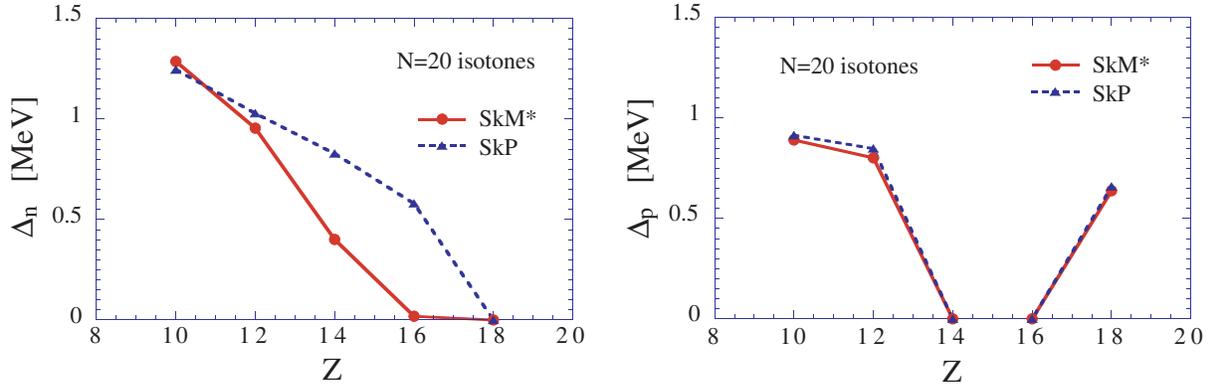}
  \end{center}
  \caption{The neutron and proton pairing gaps in N=20 isotones
  calculated in HFB with SkM* and SkP.}
  \label{FIG_DELNP}
\end{figure}

Fig.\ref{FIG_DELNP} shows the HFB neutron and proton pairing gaps
in N=20 isotones calculated with SkM* and SkP.
The pairing strengths are adjusted so as to reproduce the experimental
neutron pairing gap in $^{30}$Ne.
As the proton number increases, the neutron pairing gaps
decrease monotonically
and eventually, the neutron pairing gap becomes zero (for both SkM* and SkP)
in $^{38}$Ar as expected in stable N=20 nuclei. The interesting point is
that the N=20 shell gap
itself changes very moderately but the calculated neutron pairing gap
changes considerably from 1.26 MeV in $^{30}$Ne to zero in $^{38}$Ar.
The mechanism can be understood by the increase of the level density in the
$fp$ shell when the proton number decreases, as noted above.   
Since the neutron pairing gap 
is adjusted in $^{30}$Ne it remains close in $^{32}$Mg for both interactions,  
but large differences are seen in $^{34}$Si and $^{36}$S.
On the other hand, the calculated proton pairing 
gaps do not depend on the Skyrme force.

\begin{figure}
  \begin{center}
    \includegraphics[width=\FSA,clip]{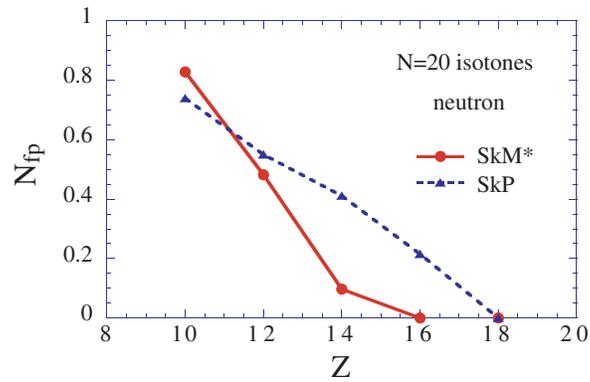}
  \end{center}
  \caption{Average number of neutrons in the $fp$ shell 
  in N=20 isotones calculated in HFB with SkM* and SkP.}
  \label{FIG_Nfp}
\end{figure}

Fig.\ref{FIG_Nfp} shows the average number of neutrons $N_{fp}$
in the $fp$ shell in N=20 isotones, 
calculated in HFB with SkM* and SkP.
According to the change of the neutron pairing gaps, $N_{fp}$ decreases  
monotonically 
from $\simeq0.8$ in $^{30}$Ne to $\simeq0.5$ in $^{32}$Mg. 
These values are very different from the prediction of the "island 
of inversion", $N_{fp}=2$ \cite{WB90} and Monte Carlo shell model, 
$N_{fp}\ge 2$ in $^{30}$Ne and $^{32}$Mg \cite{UO99}.

\section{B(E2) values and excitation energies} \label{SEC-BE2}

We have calculated the first \twop states in N=20 isotones 
in HFB-QRPA with Skyrme interactions, assuming spherical symmetry. 
At the mean-field level the ground states in N=20 isotones 
including $^{32}$Mg and $^{30}$Ne have been found to be spherical 
(see, e.g.,, \cite{TF97,RD99}). Our aim is to investigate whether 
these \twop states can be described as vibrational states
built on the spherical ground states.

\begin{figure}
  \begin{center}
    \includegraphics[width=\FSB,clip]{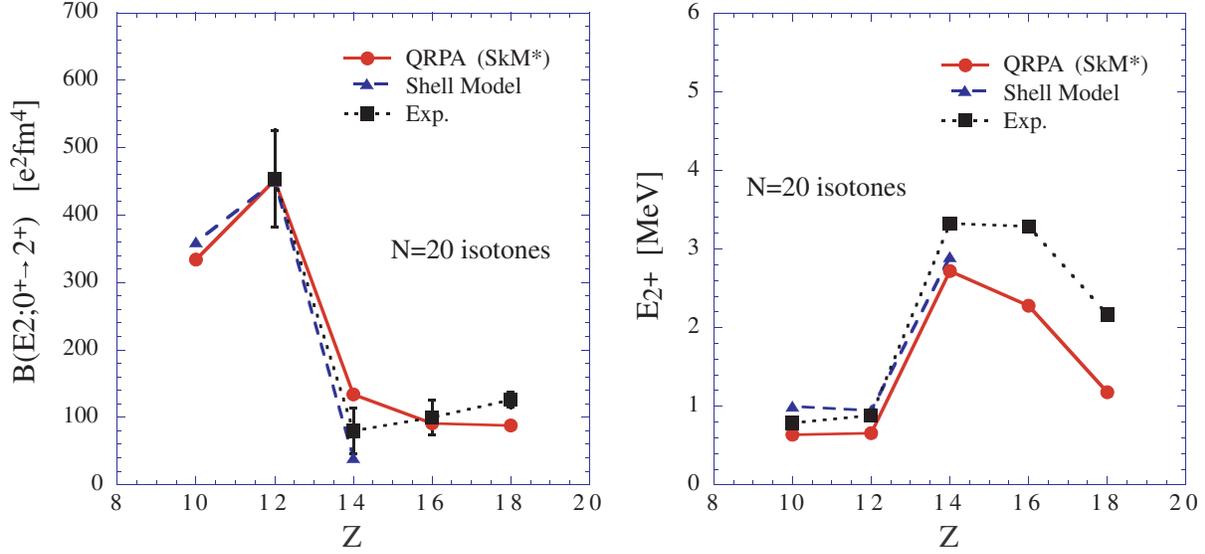}
  \end{center}
  \caption{The \betwo transition probabilities and excitation energies 
   of the first 2$^{+}$ states in N=20 isotones calculated in 
   QRPA with SkM*. For comparison the predictions of  
   MCSM\cite{UO99} and 
   the available experimental data \cite{MI95, GD84, YN02, PG99, CG01, RN89}
   are shown.}
  \label{FIG_COMP}
\end{figure}

In Fig.\ref{FIG_COMP} our results of QRPA calculations with SkM* 
are compared with the predictions of the Monte Carlo shell model (MCSM)
\cite{UO99}, and the available experimental data \cite{MI95,GD84,YN02,PG99,CG01, 
 RN89}. Our QRPA results are in good agreement with the experimental
data and they are 
consistent with the MCSM prediction of the B(E2) value 
in $^{30}$Ne which is not yet measured experimentally.
The QRPA calculations have been done with the SkM* parameter set and the fixed 
pairing strength $V_0 = -418$ MeV fm$^{-3}$ the choice of which is explained
in subsect. 2.2 .
The general properties of the first \twop states in N=20 isotones, 
namely very large B(E2) values and very low excitation energies 
in $^{32}$Mg and $^{30}$Ne are well reproduced within a single framework. 

\begin{figure}
  \begin{center}
    \includegraphics[width=\FSB,clip]{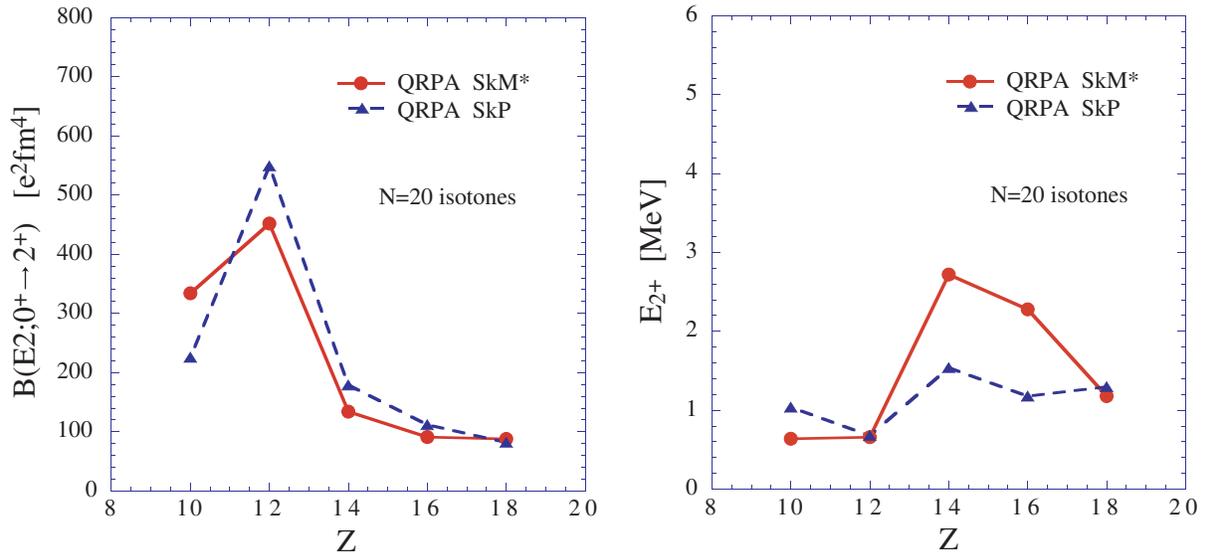}
  \end{center}
  \caption{The \betwo transition probabilities and excitation energies
    of the first 2$^{+}$ states calculated in QRPA with SkM* and 
    SkP interactions.}  
  \label{FIG_QRPA}
\end{figure}

To check the interaction dependence we have carried out QRPA calculations 
with the SkP interaction.  
Fig.\ref{FIG_QRPA} shows the B(E2) values and excitation 
energies of the first \twop states with SkM* and SkP.
Concerning the B(E2) values we get similar results, especially very large B(E2) 
values in $^{32}$Mg and $^{30}$Ne. On the other hand, large differences are seen 
in the excitation energies in $^{34}$Si and $^{36}$S. 
This can be understood by the difference in the neutron pairing 
correlations shown in Fig.\ref{FIG_DELNP}.
In $^{30}$Ne, $^{32}$Mg and $^{38}$Ar the neutron pairing gaps calculated 
with SkM* and SkP are almost the same 
while they are somewhat different in $^{34}$Si 
and $^{36}$S. Because the neutron pairing gaps are larger in SkP than in SkM*, 
the excitation energies become lower with SkP than with SkM*. 

\begin{figure}
  \begin{center}
    \includegraphics[width=\FSB,clip]{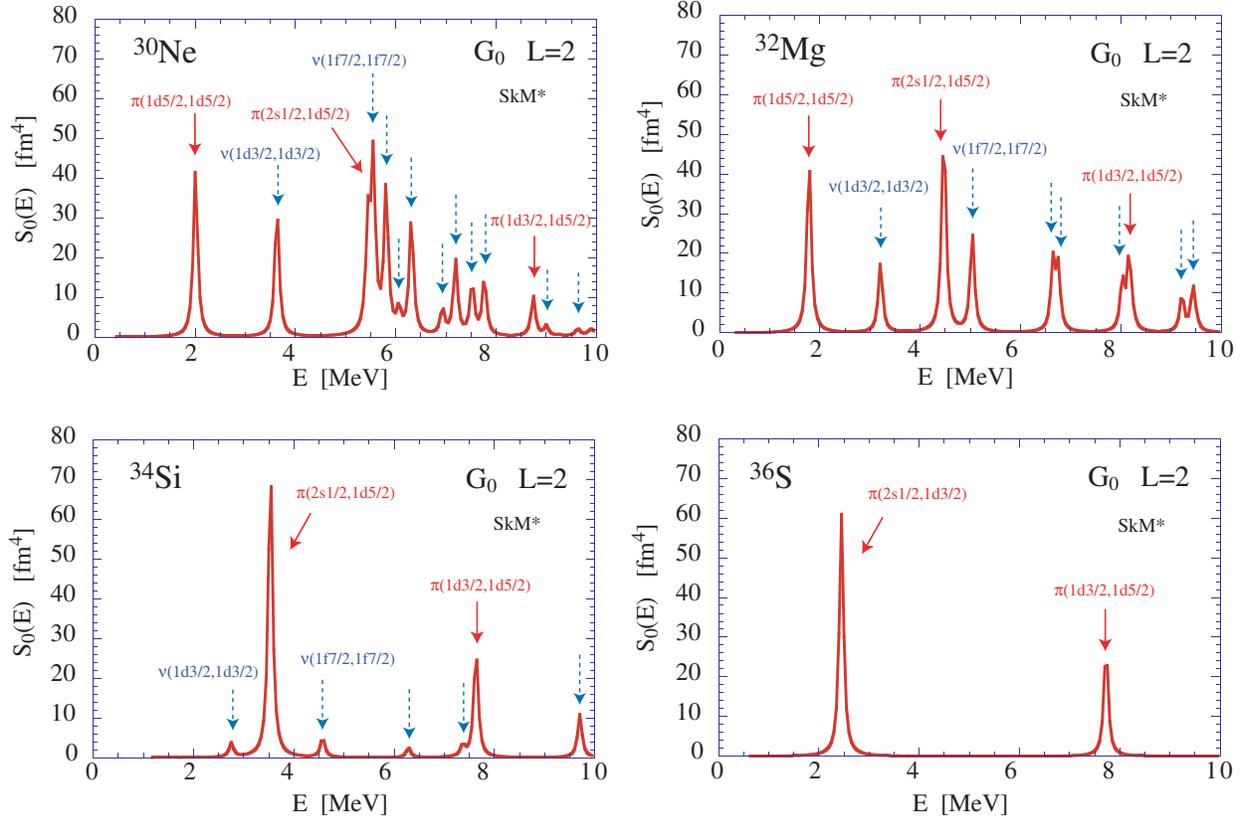}
  \end{center}
  \caption{The unperturbed isoscalar quadrupole strength functions 
           in N=20 isotones calculated with SkM*.
           The peaks indicated by straight (dotted) arrows correspond to 
           proton (neutron) two-quasiparticle configurations.}
  \label{FIG_G0}
\end{figure}

\begin{figure}
  \begin{center}
    \includegraphics[width=\FSB,clip]{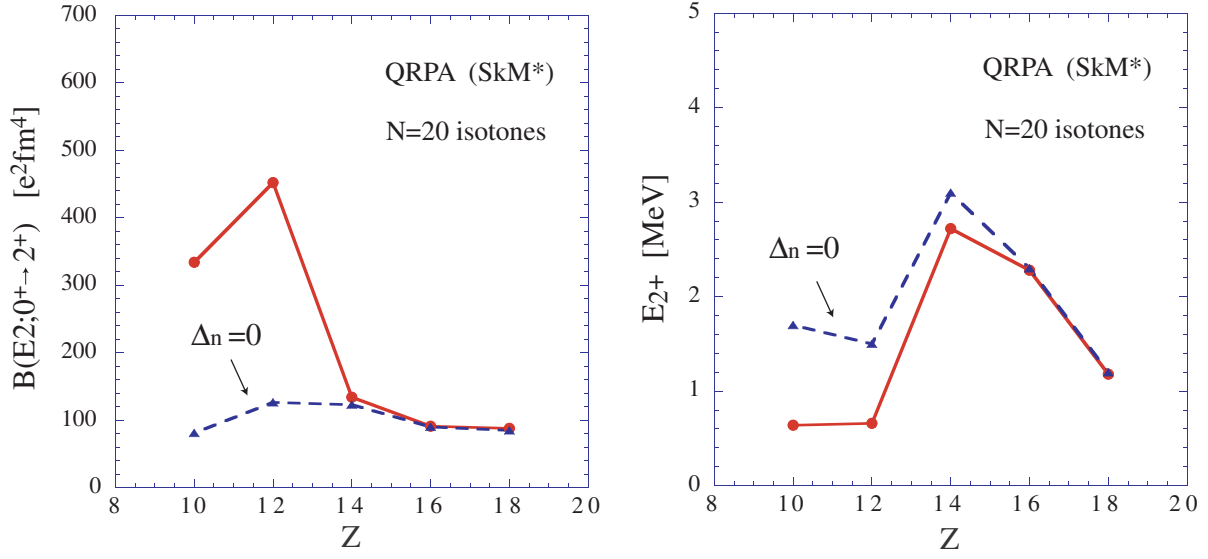}
  \end{center}
  \caption{The \betwo values and the excitation energies of the first 
  \twop states in N=20 isotones calculated with/without neutron pairing 
  correlations. Proton pairing is included in both cases.}
  \label{FIG_QRPA_V000}
\end{figure}

We now explain how the neutron pairing correlations are important
to make the B(E2) values larger and the excitation energies lower.
To see which two-quasiparticle configurations contribute to make the
low-lying \twop states, we show the unperturbed isoscalar quadrupole strength
functions in N=20 isotones calculated with SkM* in Fig.\ref{FIG_G0}.
The peaks indicated by straight (dotted) arrows correspond to
proton (neutron) two-quasiparticle configurations.
All these neutron two-quasiparticle configurations appear because of
the neutron pairing correlations. Many peaks of the neutron configurations
are seen in $^{30}$Ne, $^{32}$Mg. On the other hand, the neutron configurations
are negligible in $^{34}$Si and they completely disappear in $^{36}$S.
The B(E2) values are primarily made of the proton configurations in the
$sd$-shell 
but the neutron 
configurations assist to make the B(E2) values 
larger by coherence between protons and neutrons.
Actually, if the neutron pairing correlations are neglected the B(E2) values become 
very small and the excitation energies are sizably higher 
in $^{32}$Mg and $^{30}$Ne, 
as shown in Fig.\ref{FIG_QRPA_V000}.
Under these considerations, we can conclude that the very large B(E2) values and 
the very low excitation energies in $^{32}$Mg and $^{30}$Ne appear thanks to 
the presence of the neutron pairing correlations. 
We have seen in sect. \ref{SEC-GS} that, around the drip line the origin of neutron
pairing correlations lies in 
the different behavior of the single-particle levels with different orbital 
angular momentum {\it l} as the levels approach the separation threshold.

Generally speaking, neutron 2p-2h configurations across N=20 
can originate 
from deformation effects or pairing effects. Both effects can {\it a priori}
contribute in the nucleus $^{32}$Mg. 
We have shown that a spherical QRPA description, i.e., putting emphasis on
the pairing aspects and neglecting the possible deformation effects, can
give very satisfactory results.  
In the previous studies based on shell model calculations\cite{FO92,CN98,UO99} 
the importance of neutron 2p-2h configurations for describing the B(E2)
values 
in $^{32}$Mg, $^{30}$Ne was shown, 
but the respective roles of pairing and deformation were not clear. 
The angular momentum projected GCM calculations\cite{RE02,RE03} 
are successful in reproducing  
the general trend of the B(E2) values and the excitation energies in Mg and Ne 
isotopes. However, their predicted excitation energies are somewhat higher
than experiment. 
This may be due to the weakness of neutron pairing correlations in these
calculations. 

\begin{figure}
  \begin{center}
    \includegraphics[width=\FSA,clip]{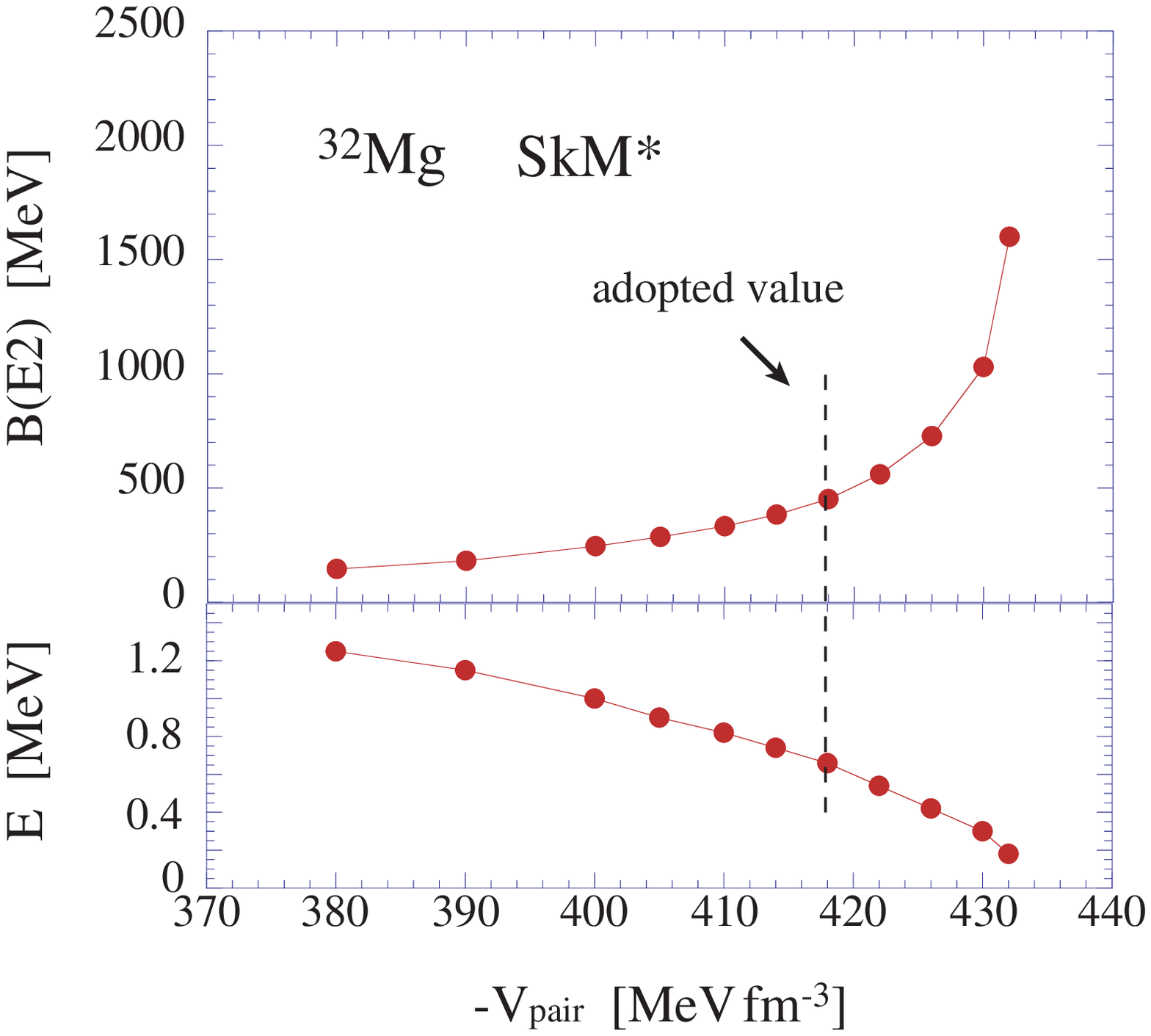}
  \end{center}
  \caption{The B(E2) value and excitation energy of 
      the first 2$^{+}$ state in $^{32}$Mg calculated 
      in QRPA with SkM*, as a function of the pairing strength $V_{pair}$.}
  \label{FIG_V-dep}
\end{figure}

Before closing this section we would like to make a brief comment on the stability of 
QRPA solutions. As it is well known, 
when a QRPA eigenvalue is approaching zero 
the solution suffers instability and the transition probability diverges. 
Since our QRPA solutions for $^{32}$Mg and $^{30}$Ne have very low energies, 
we have to check whether the calculated B(E2) values are really meaningful
or just 
spurious results. Fig.\ref{FIG_V-dep} shows the pairing strength $V_{pair}$ dependence 
of the excitation energy and the B(E2) value in $^{32}$Mg. 
If $|V_{pair}|$ increases, the excitation energy decreases and the B(E2) value 
increases. This behavior is the result of competition between two effects. 
First, 
the pairing gap and also 
the quasiparticle energies increase with increasing $|V_{pair}|$. 
Therefore, the two-quasiparticle energies 
and the QRPA excitation energies should increase and the B(E2) values should
decrease.  
Second, if $|V_{pair}|$ increases, many two-quasiparticle configurations 
can contribute to make the \twop state and the collectivity increases. 
In this case the QRPA excitation energies would decrease and the B(E2) values  
would increase. In the $^{32}$Mg case the second effect wins (cf.
Fig.\ref{FIG_V-dep}).
As $|V_{pair}|$ is increasing, the excitation energy becomes 
lower and the B(E2) value
increases linearly up to $|V_{pair}| \simeq 422$ MeV fm$^{-3}$.  
Above $|V_{pair}| \simeq 426$ MeV fm$^{-3}$ the B(E2) value starts to 
diverge.
Because our adopted pairing strength is $V_{pair} = - 418$ MeV fm$^{-3}$, 
we confirm that our QRPA solution is meaningful.


\section{Conclusion}  \label{SEC-CONC}

We have studied the first \twop states in N=20 isotones by 
the HFB-QRPA model 
with Skyrme interactions. 
The residual interaction is consistently derived from the Skyrme Hamiltonian, 
keeping all its explicit momentum dependence.

Because of the different behaviors of the neutron $1f_{7/2}$, $2p_{3/2}$ 
and $2p_{1/2}$ levels when the single-particle energies are approaching zero, 
the neutron pairing gaps have 
finite values. This mechanism breaks the N=20 magicity in $^{32}$Mg and $^{30}$Ne.

Within our consistent QRPA calculation with spherical symmetry 
the \betwo values and the excitation energies of the first 
\twop states in N=20 isotones including $^{32}$Mg and $^{30}$Ne are well described. 
The existing experimental data are reproduced quantitatively.
The B(E2) value in $^{30}$Ne has not been measured yet but the QRPA value   
is consistent with the prediction of the MCSM.
The important role of the neutron pairing correlation is emphasized.
If the neutron pairing is dropped, we cannot get the correct B(E2) value 
and excitation energy in $^{32}$Mg and $^{30}$Ne.
In the real $^{32}$Mg nucleus, both neutron pairing and deformation effects  
may coexist and help to make the large B(E2) value, but our calculation 
shows that neutron pairing correlations are essential.


\section{ACKNOWLEDGMENTS}

We acknowledge 
Prof. K. Matsuyanagi for valuable discussions.
We thank 
Prof. T. Suzuki (Fukui) and 
Prof. I. Hamamoto 
for useful comments.  
We also thank 
Dr. E. Khan for many discussions. One of the authors (MY) is very grateful
to the members of the Theory Group of IPN-Orsay for their hospitality during
his stay. 
Numerical computation in this work was carried out at the 
Yukawa Institute Computer Facility.

%
%

\newpage

%
%

\begin{thebibliography}{0}



\bibitem{TK75}
C. Thibault, R. Klapisch, C. Rigaud, A. M. Poskanzer, R. Prieels, L. Lessard, 
W. Reisdorf, 
Phys. Rev. C12 (1975) 644.


\bibitem{MI95}
T. Motobayashi, Y. Ikeda, Y. Ando, K. Ieki, M. Inoue, N. Iwasa, T. Kikuchi, 
M. Kurokawa, S. Moriya, S. Ogawa, H. Murakami, S. Shimoura, Y. Yanagisawa, 
T. Nakamura, Y. Watanabe, M. Ishihara, T. Teranishi, H. Okuno, R. F. Casten 
Phys. Lett. 346B (1995) 9.




\bibitem{GD84}
D. Guillemaud-Mueller, C. Detraz, M. Langevin, F. Naulin, 
M. De Saint-Simon, C. Thibault, F. Touchard, M. Epherre, 
Nucl. Phys. A426 (1984) 37. 

\bibitem{YN02}
Y. Yanagisawa, M. Notani, H. Sakurai, S. Shimoura, H. Iwasaki, S. Michimasa, 
K. Ue, M. Kurokawa, N. Iwasa, M. Kunibu, H. Baba, T. Gomi, A. Saito, T. Minemura, 
Y. U. Matsuyama, Y. Higurashi, S. Kanno, M. Serata, E. Takeshita, S. Takeuchi, 
K. Yamada, K. Demichi, H. Hasegawa, H. Akiyoshi, N. Fukuda, K. Yoneda, N. Aoi, 
N. Imai, T. Sugimoto, T. Nakamura, T. Motobayashi, 
RIKEN Accelerator Progress Report 2001 (2002) 71. 

\bibitem{CF75}
X. Campi, H. Flocard, A. K. Kerman, S. Koonin, 
Nucl.Phys. A251 (1975) 193. 

\bibitem{WC80}
B. H. Wildenthal, W. Chung, 
Phys.Rev. C22 (1980) 2260.

\bibitem{WS81}
A. Watt, R. P. Singhal, M. H. Storm, R. R. Whitehead, 
J. Phys. G7 (1981) L145.

\bibitem{PR87}
A. Poves, J. Retamosa, 
Phys. Lett. 184B (1987) 311.

\bibitem{WB90}
E. K. Warburton, J. A. Becker, B. A. Brown, 
Phys. Rev. C41 (1990) 1147.

\bibitem{FO92}
N. Fukunishi, T. Otsuka, T. Sebe, 
Phys. Lett. 296B (1992) 279.

\bibitem{CN98}
E. Caurier, F. Nowacki, A. Poves, J. Retamosa, 
Phys. Rev. C58 (1998) 2033.

\bibitem{UO99}
Y. Utsuno, T. Otsuka, T. Mizusaki, M. Honma, 
Phys. Rev. C60 (1999) 054315. 


\bibitem{TF97} 
J. Terasaki, H. Flocard, P. -H. Heenen, P. Bonche, 
Nucl. Phys. A621 (1997) 706.

\bibitem{RD99}
P. -G. Reinhard, D. J. Dean, W. Nazarewicz, J. Dobaczewski, J. A. Maruhn, M. R. Strayer, 
Phys. Rev. C60 (1999) 014316.


\bibitem{PG00}
S. P$\acute{\mbox{e}}$ru, M. Girod, J. F. Berger, 
Eur. Phys. J. A 9 (2000) 35.


\bibitem{RE02}
R. Rodriguez-Guzman, J. L. Egido, L. M. Robledo, 
Nucl. Phys. A709 (2002) 201.

\bibitem{RE03}
R. Rodriguez-Guzman, J. L. Egido, L. M. Robledo, 
Eur. Phys. J. A 17 (2003) 37. 

\bibitem{KH02}
M. Kimura, H. Horiuchi, 
Prog. Theor. Phys. 107 (2002) 33. 


\bibitem{YS01}
K. Yoneda, H. Sakurai, T. Gomi, T. Motobayashi, N. Aoi, N. Fukuda, U. Futakami, 
Z. Gacsi, Y. Higurashi, N. Imai, N. Iwasa, H. Iwasaki, T. Kubo, M. Kunibu, 
M. Kurokawa, Z. Liu, T. Minemura, A. Saito, M. Serata, S. Shimoura, S. Takeuchi, 
Y. X. Watanabe, K. Yamada, Y. Yanagisawa, K. Yogo, A. Yoshida, M. Ishihara, 
Phys. Lett. 499B (2001) 233.

\bibitem{Gu02}
D. Guillemaud-Mueller, 
Eur. Phys. J. A 13 (2002) 63.

\bibitem{YK03}
M. Yamagami, E. Khan, Nguyen Van Giai,
to be published. 

\bibitem{KS02}
E. Khan, N. Sandulescu, M. Grasso, Nguyen Van Giai, 
Phys. Rev. C66 (2002) 024309.

\bibitem{RM87}
S. Raman, C. H. Malarkey, W. T. Milner, C. W. Nestor, Jr., P. H. Stelson, 
At. Data Nucl. Data Tables 36 (1987) 1.


\bibitem{GB94} B. Gall, P. Bonche, J. Dobaczewski, H. Flocard, P. -H. Heenen, 
Z. Phys. A348 (1994) 183.

\bibitem{TH95} J. Terasaki, P. -H. Heenen, P. Bonche, J. Dobaczewski, H. Flocard,
Nucl. Phys. A593 (1995) 1. 

\bibitem{YM01} M. Yamagami, K. Matsuyanagi, M. Matsuo, 
Nucl. Phys. A693(2001) 579. 

\bibitem{BQ82} J. Bartel, P. Quentin, M. Brack, C. Guet, H. -B. Hakansson, 
Nucl. Phys. A386 (1982) 79. 

\bibitem{DF84} J. Dobaczewski, H. Flocard, J. Treiner, 
Nucl. Phys. A422 (1984) 103.



\bibitem{SD98} W. Satula, J. Dobaczewski, W. Nazarewicz, 
Phys. Rev. Lett. 81 (1998) 3599. 

\bibitem{Ma01}
M. Matsuo, 
Nucl. Phys. A696 (2001) 371 

\bibitem{HL01}
I. Hamamoto, S. V. Lukyanov, X. Z. Zhang, 
Nucl. Phys. A683 (2001) 255.

\bibitem{PG99}
B. V. Pritychenko, T. Glasmacher, P. D. Cottle, M. Fauerbach, R. W. Ibbotson, 
K. W. Kemper, V. Maddalena, A. Navin, R. Ronningen, A. Sakharuk, H. Scheit, 
V. G. Zelevinsky 
Phys. Lett. 461B (1999) 322 ; Erratum Phys.Lett. 467B (1999) 309. 

\bibitem{CG01}
V. Chiste, A. Gillibert, A. Lepine-Szily, N. Alamanos, F. Auger, J. Barrette, 
F. Braga, M. D. Cortina-Gil, Z. Dlouhy, V. Lapoux, M. Lewitowicz, R. Lichtenthaler, 
R. Liguori Neto, S. M. Lukyanov, M. MacCormick, F. Marie, W. Mittig, F. de Oliveira Santos, 
N. A. Orr, A. N. Ostrowski, S. Ottini, A. Pakou, Yu. E. Penionzhkevich, P. Roussel-Chomaz, J.L.Sida 
Phys. Lett. 514B (2001) 233.







\bibitem{RN89}
S. Raman, C. W. Nestor, Jr., S. Kahane, K. H. Bhatt, 
At. Data Nucl. Data Tables 42 (1989) 1. 


\end{thebibliography}
\end{document}